\documentclass[fleqn,usenatbib]{mnras}

\usepackage{newtxtext,newtxmath}

\usepackage[T1]{fontenc}
\usepackage{ae,aecompl}

\usepackage{graphicx}	
\usepackage{amsmath}	
\usepackage{amssymb}	

\newcommand{\nn}{\mbox{} \nonumber \\ \mbox{} }
\newcommand{\ba}{\begin{eqnarray}}
\newcommand{\ea}{\end{eqnarray}}
\newcommand{\om}{\omega}

\newcommand{\Bf}{{magnetic field}}

\newcommand{\Bfs}{{magnetic fields}}
\newcommand{\NS}{neutron star}

\newcommand\eg{\textit{e.g.}}

\newcommand\co{\textit{cocoon}}
\newcommand\sj{\textit{structured jet}}
\newcommand\fjc{\textit{fast jet -- cocoon}}

\def\be{\begin{equation}}
\def\ee{\end{equation}}

\title[Second peak  in the afterglow of GW170817]{Prediction of the second peak in the afterglow of  GW170817}

\author[M. V. Barkov et al.]{
Maxim V. Barkov,$^{1,2}$\thanks{E-mail: mbarkov@purdue.edu (BMV)}
Adithan Kathirgamaraju,$^{1}$
Yonggang Luo,$^{1}$
\newauthor Maxim Lyutikov,$^{1}$
 and Dimitrios Giannios,$^{1}$\\
$^{1}$ Department of Physics and Astronomy, Purdue University, West Lafayette, IN 47907-2036, USA\\
$^{2}$ Astrophysical Big Bang Laboratory, RIKEN, 351-0198 Saitama, Japan }

\date{Accepted XXX. Received YYY; in original form ZZZ}

\pubyear{2018}

\begin{document}
\label{firstpage}
\pagerange{\pageref{firstpage}--\pageref{lastpage}}
\maketitle

\begin{abstract}
We performed calculations of the late radio and X-ray afterglow of GRB/GW170817 in the cocoon-jet paradigm, predicting  appearance of a second peak  in the afterglow light curve  $\sim$ one-three years after the explosion. The model  assumes  that the prompt emission and early afterglows originate from a cocoon generated during break-out of the delayed magnetically powered jet. As the jet breaks out from the torus-generated wind, a nearly isotropic mildly relativistic outflow is generated; at the same time the primary jet accelerates to high Lorentz factors and avoids detection. As the fast jet slows down, it should become visible to the off-axis observer. Thus, the model has a clear prediction: the X-ray and radio afterglows should first experience a decay, as the cocoon slows down, followed by a rebrightening when the primary jet  starts emitting toward an observer.
\end{abstract}




\section{Introduction}
On 2017 August 17, LIGO and Virgo detector discover the gravitational-wave (GW) transient
GW170817~\citep{GW-trigger}, which was consistent with the coalescence of a binary neutron star system. 
Two seconds later GRB~170817A was registered by GBM/\textit{Fermi} \citep{gol17} and  SPI-ACS/INTEGRAL
\citep{sav17}  experiments. A the optical counterpart was detected by a large number of ground-based facilities \citep{2017ApJ...848L..13A}.

GRB/GW170817 was unusual in many respects.  The prompt gamma-ray emission consisted of two distinctive components -  a hard short pulse delayed by $\sim 2$ seconds with respect to the  LIGO signal followed by a weaker, softer
thermal pulse with $T\sim 10 $~keV lasting for another $\sim2$ seconds, (see Fig.~1 in \cite{2018ApJ...852L..30P}).  The appearance of a thermal component at
the end of the burst is unusual for short GBRs.  Both the hard and the soft components do not satisfy the Amati relation, making GRB~170817A distinctively different  from other short GRBs \citep{2018ApJ...852L..30P,2018MNRAS.475.2971B}.

\section{Model}

\begin{figure}
\includegraphics[width=\columnwidth]{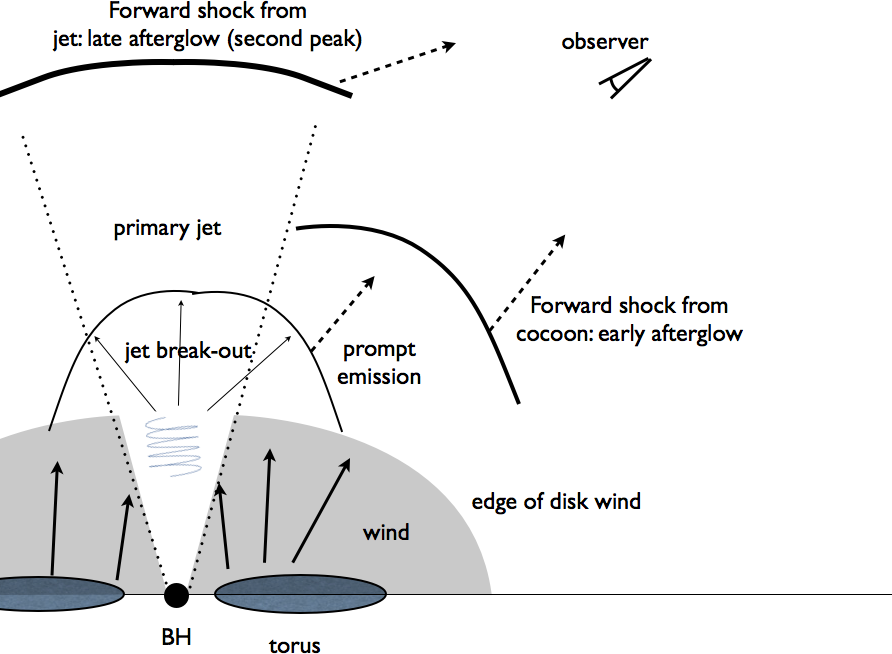} 
\caption{Cartoon of the model. Hot torus (left from the disrupted neutron stars) creates a dense, mildly relativistic wind (shaded region).  After $\sim 1$ second, when enough magnetic flux is accumulated on the BH, the BH launches a Blandford-Znajek-powered jet. After the jet reaches the edge of the confining wind its head part experiences a break-out, creating a nearly spherical outflow;  this generates  the  prompt emission.  Later on,
the interaction of this now nearly-spherically expanding part of  the jet  with the surrounding medium generates the forward shock - this leads to the production of early   the early afterglow that has been observed so far.  Most of the jet accelerates to high Lorentz factors; the radiation from the corresponding forward shock is beamed away from the observer. Only after the jet-driven forward shock decelerated it will become visible, and should  generate the late bump in the afterglow.}
\label{fig:sk}
\end{figure}

The detection of the EM signal contemporaneous with gravitational waves is  consistent with the binary NS scenario for short GRB \citep{bnpp84,p86,1989Natur.340..126E}.
Currently, there are several competing models for the prompt and afterglow emission from GW170817: (i) radially stratified quasi-spherical ejecta (\co) traveling at mildly relativistic speeds 
(e.g., \citealt{mooley2018,kasliwal2017,gnp17}); (ii) emission from off-axis collimated  ejecta characterized by a narrow cone of ultra-relativistic material 
with slower wings extending to larger angles (\sj) (e.g., \citealt{lamb2018,troja2018,davanzo2018,xie2018,2018arXiv180103531M,kbg17,lazzati2017}). In the \sj{} scenario, the GW170817 merger powered a normal SGRB directed away from the line of sight; (iii) \fjc{} model - we describe it next in more detail.

The key points of the  \fjc{}  model is described in \cite[][]{2018ApJ...852L..30P}; also  see Fig~\ref{fig:sk}. An active stage of a merger lasts $\sim 10-100$ milliseconds after which the {\NS}s collapse into BH. During the merger  an accretion torus of  $\sim 0.1 M_{\odot}$ forms around the BH with a viscous time  $\sim$0.1~s \citep[\eg][]{rlps16,2017ApJ...850L..37P}. At the same time, \Bfs\ are amplified within the disk to $\sim 10^{15}$~G \citep{2011ApJ...732L...6R,2012PhRvD..86j4035L,rlps16} due to  the development of  MRI and the presence of the velocity shear.
Qualitatively, as the matter is accreted onto the BH, the BH accumulates magnetic flux. At the same time baryons
slide off into the BH along \Bf\ lines, leaving  polar  regions with low density. This creates conditions favorable for the operation of the  Blandford-Znajek (BZ) mechanism \citep{bz77}. As a result, the  accumulation of the magnetic flux leads to a delay for the jet to switch-on. 
The  BZ jet then propagates through a pre-existing  dense wind with mildly relativistic velocity \citep{2018ApJ...852L..30P,2018MNRAS.475.2971B}. As it breaks out from the wind, it generates a nearly isotropic cocoon, Fig.~\ref{fig:sk}.  The wind emission shocked by the breaking-out jet produces the soft tail. After the  break-out the primary jet accelerates and becomes invisible to the observer.  The inclination of the binary system come directly from the GWs signal $\theta_{\rm obs}\approx30^o\pm 10^o$ \citep{2017ApJ...848L..13A,2018arXiv180404179F}.

Observationally,  Chandra and VLA  \citep{2017Natur.551...71T,2018arXiv180106516T,2018arXiv180103531M,2018arXiv180502870A} show that  GRB/GW170817  was steadily 
brightening with time, and  have now reached its peak and starts to decay. The very simple power-law spectrum was extending for eight orders of magnitude in frequency. The measurement of the power low index $p=2.17$ 
indicates that radiation should come from ejecta with $\Gamma\sim3-10$.

We interpret these observations 
as non-thermal synchrotron emission coming  from the ``break-out'' (nearly spherical) part of the mildly relativistic  forward shock. But 
observations up to $t\le 250$ days  have not been able to distinguish the above scenarios, because of the observed emission will be dominated by radiation from 
mildly relativistic material \citep{2018arXiv180103531M} present in all the models.

Most importantly, at later times  {\it  the  models predict qualitatively different behavior.}
Both the \co{} and \sj{} models should produce  only one bump in X-ray light curve by  the jet or the  cocoon. 
On the other hand, the \fjc{} model  \citep{2018ApJ...852L..30P} has two active components - a cocoon formed during  
jet breakout and  an ultra-relativistic jet. The initial rise of $x$-ray light curve (see red dots Figure~\ref{fig:sb}) 
is formed by a cocoon - a shock break-out.  As we discuss in this paper, later on -- a few years after GW event --  the \fjc{}  model  predicts rebrightening of the afterglow as the primary jet slows down and becomes visible (see Fig.~\ref{fig:sk}).
The detection of second X-ray  or radio bump will be a smoking gun for \fjc{} model and rule out one component models. Calculations of the properties of the predicted  second afterglow bump is the key point of the paper.

\section{Results}

In this work we perform three types of calculations of the predicted second peak: (i) using  analytical estimates from \citep[][\S \ref{Analytic}]{2002ApJ...579..699N,2017arXiv171006421G}; (ii) model light curves from the  Afterglow library \citep[][\S \ref{library}]{2012ApJ...747L..30V}; (iii) in-house  numerical calculations of the synchrotron emissivity of the  relativistically expanding and synchrotron cooling plasma, \S \ref{inhouse}.

\subsection{The fiducial parameters}

The accretion torus/disc after the NS-NS merger can be relatively massive $\sim 0.1 M_{\odot}$ \citep[\eg][]{2017ApJ...850L..37P}. 
The disc  produces a dense mildly relativistic wind with mass $\sim 0.05 M_{\odot}$   \citep{bb99,mpq08}  
and also supplies the central BH with magnetic flux needed to launch the BZ jet.
The accretion rate on BH horizon can be estimated as \citep[see more details in ][]{2018ApJ...852L..30P}
\begin{equation}
\dot{M}_{\rm BH} \approx 0.002 \frac{M_{\rm d,-1}}{t^{5/3}} \; M_{\odot}/\mbox{s}\;.
\label{eq:mdot}
\end{equation}
Here $M_{\rm d} = 0.1 M_{d,-1}M_{\odot}$ is mass of the disc in solar mass units and time is in seconds. 
The BZ jet power can be \citep{bz77,BK08b,bp11} as high as
\begin{equation}
L_{\rm BZ} \approx 5\times10^{50} \frac{M_{d,-1}}{t_{0.3}^{5/3}} \; \mbox{erg}/\mbox{s}\;,
\label{eq:lbz}
\end{equation}
here we assume efficiency of BZ jet formation $C(a_{BH}) \approx a_{\rm BH}^{2.4}$ with BH spin parameter $a_{\rm BH} = 0.7$ \citep{rlps16,rgl16}.

The opening angle of the jet is unknown. The second bump can be detected if jet is relatively narrow and powerful.
Following \citep{2018ApJ...852L..30P} the opening angle of the jet can be assumed $\theta_{j} = 0.1 \approx 5^{\rm o}$, so the isotropic jet power can be as high as 
\begin{equation}
E_{\rm iso,max} \approx \frac{2}{\theta_{j}^2} \int_{t=2}^{\infty} L_{\rm BZ} dt \approx 10^{53} \mbox{ ergs}.
\label{EISO}
\end{equation}
This is the estimate of the  primary jet energy that we will use in the calculations.

Initially, the primary jet emission   is beamed away from the observer. As the jet-driven blast wave slows down it becomes visible.  In the following, we perform calculations  to address the question: What are the conditions required to produce an observable second afterglow bump from the primary  jet.

\subsection{Analytic estimates for detectability of the second peak.} \label{analyticestimate}
\label{Analytic}

\begin{figure} 
\includegraphics[width=\columnwidth]{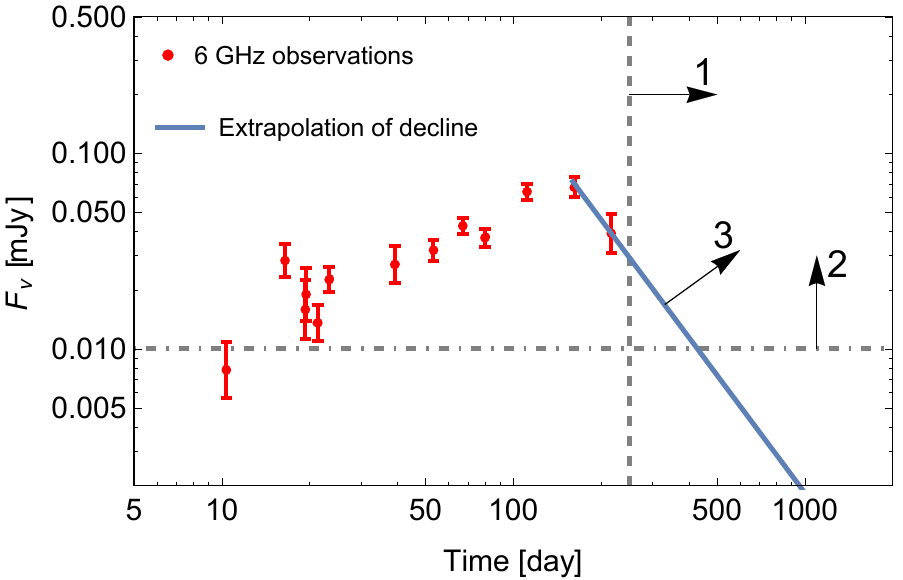}\\
\includegraphics[width=\columnwidth]{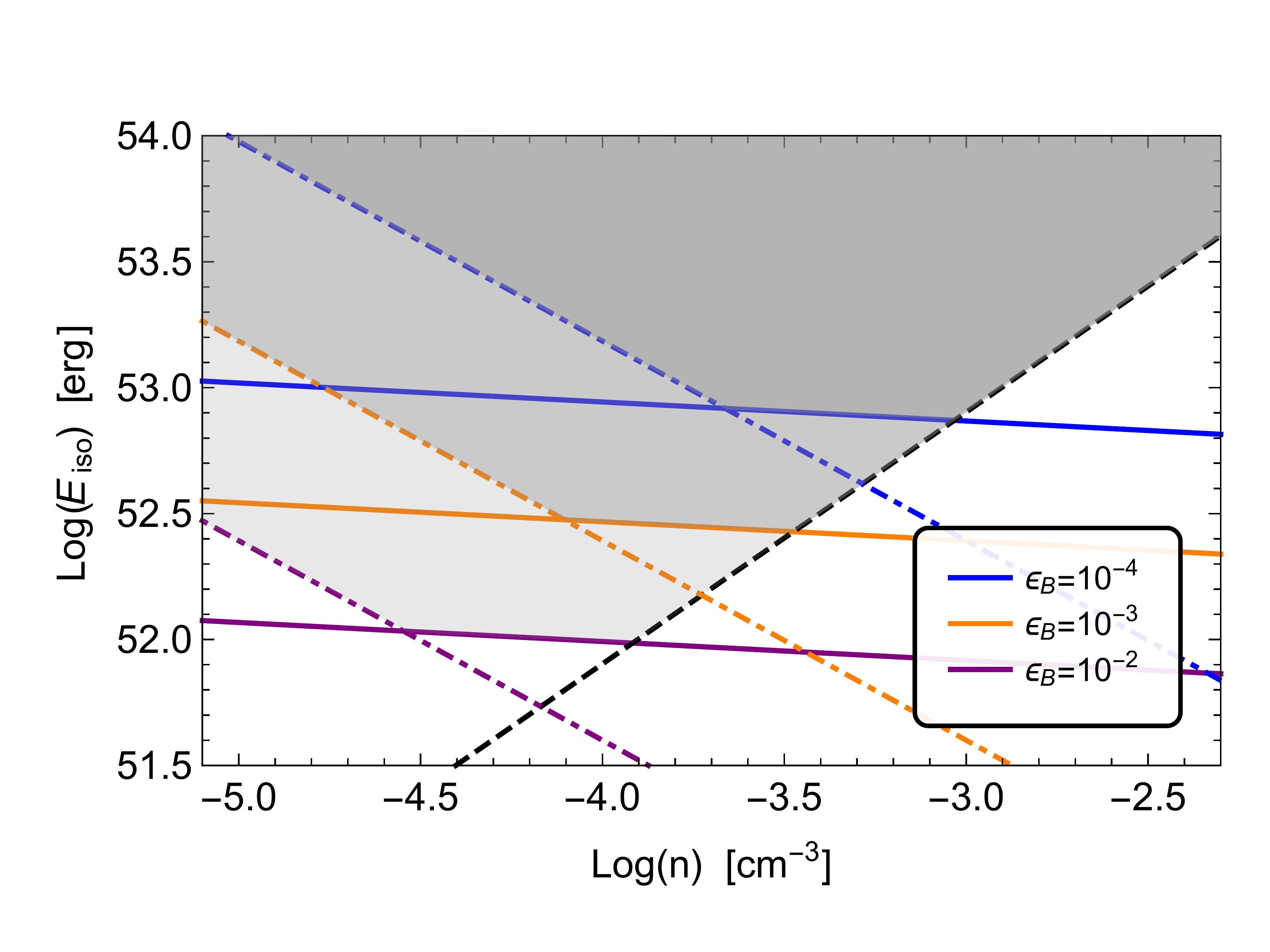}
\caption{(Top) Depiction of the three criteria required for detectability of second peak. The plot shows observed radio data (red points) and a power law extrapolation of the observed decline (solid line) 
\citep[data taken from ][]{2018arXiv180103531M,2018arXiv180502870A}.
Vertical dashed line marks a time of 250 days and dot-dashed horizontal line indicates a radio detectability limit of 10 $\mu$Jy. The numbered arrows point to the region where the second peak must lie in order to be detectable and the numbers label the criteria described in Section \ref{analyticestimate}. 
In short, 1) requires peak time to be greater than 250d, 2) requires peak flux be above detector sensitivity and 3) requires peak flux be larger than  cocoon emission. (Bottom) A figure exploring the parameter space in isotropic equivalent energy of the jet ($E_{\rm iso}$) vs. external density ($n$). Shaded regions mark the parameter space where the second peak will be detectable (where all 3 criteria mention in top panel and Section \ref{analyticestimate} are satisfied). 
Regions above dashed black line satisfy criterion 1, above dot-dashed lines satisfy criterion 2 and above solid lines satisfy criterion 3. Colors indicate the value of $\epsilon_{\rm B}$ used for the solid and dot-dashed lines. 
See section \ref{analyticestimate} for analytic expressions of the lines, 
shaded regions and for relevant parameters used.
}
\label{paramspace}
\end{figure}

Let us first obtain  simple analytic constraints on parameters in order for the second peak associated with the afterglow of the jet to be detectable. We consider the second peak to be detectable if the following 3 criteria are satisfied. 
1) The time of the second peak ($t_{\rm peak}$) must be greater than the time of current observations ($\sim 250$~d), otherwise this peak would have been already detected or can be weaker than the cocoon component, in which case the second peak will not be detectable. 
2) The peak flux must be greater than the sensitivity limit of the detector (for radio at 6 GHz we use a limit of 10 $\mu$Jy). 
3) The value of the peak flux must be larger than that of the cocoon at the time of the peak. To estimate the flux from the cocoon component at a late time, we use a power law extrapolation of current observations. 
These criteria are depicted in the top panel of Fig.~\ref{paramspace} where the numbered arrows show the regions where the corresponding criteria mentioned above are satisfied. In this figure we use the radio data at 6GHz as an example, the dashed vertical line marks a time of 250 days, the dot-dashed horizontal line marks a detectability limit of 10 $\mu$Jy, and the solid line shows the extrapolation of the decline in the observed emission assuming it is $\propto t^{-2}$ as estimated in \cite{2018arXiv180502870A}. 
Below we always assume the observing frequency $\nu_{\rm m}<\nu<\nu_{\rm c}$ and $(1+z)\approx 1$ which is valid for GW170817.

The peak in the afterglow occurs as the beaming angle of the emission from the core of the jet increases and reaches the line of sight of the observer ($1/\Gamma\sim\theta_{\rm obs}$). This peak occurs at a time \citep[e.g.,][]{2017arXiv171006421G} 

\begin{equation}
t_{\rm peak}\approx 280 \left(\frac{E_{\rm iso,52}}{n_{-4}}\right)^{\frac{1}{3}}\left(\frac{\theta_{\rm obs}}{25^{\circ}}\right)^{\frac{8}{3}} \mbox{ days}.
\label{tpeak}
\end{equation}
Criterion 1) requires $t_{\rm peak}\gtrsim 250 d$, using the above equation and assuming $\theta_{\rm obs}=25^{\circ}$, we get $E_{\rm iso,52}\gtrsim 0.8 \,n_{-4}$. This inequality is satisfied by all the regions above the dashed black line in the bottom panel of Fig. \ref{paramspace} which shows $E_{\rm iso}$ vs $n$.

The peak flux of the afterglow can be obtained by substituting $t_{\rm peak}$ in analytic expressions for the  flux for post jet break light curves 
\citep[e.g.,][]{2002ApJ...579..699N,2017arXiv171006421G}
$$
F_{\rm peak}\approx C(p)\left(1-{\rm\cos}\theta_j\right)^{\frac{p+3}{3}}\times
$$
\begin{equation}
\qquad \qquad D_{\rm L,26}^{-2}\,\epsilon_{\rm e,-1}^{p-1}\epsilon_{B,-2}^{\frac{p+1}{4}}\,n^{\frac{1+p}{4}}\,\theta_{\rm obs}^{-\frac{8p}{3}}\,E_{\rm iso,52}\,\nu_{9.7}^{\frac{1-p}{2}} \;{\rm mJy},
\end{equation}
Where $\theta_{\rm obs}$ is in degrees and 
$$
C(p)\approx7440(p-0.04)\left(\frac{p-2}{p-1}\right)^{p-1}\left(1.13\times 10^{-20}\right)^{-p}10^{-14.96p}
$$
These analytic estimates agree within a factor $\sim 1.5$ when compared to the light curves shown in Fig. \ref{fig:sb}.
Criterion 2 requires $F_{\rm peak}$ to be larger than the detector sensitivity, in the case of radio we use $F_{\rm peak}\gtrsim 0.01$~mJy. Substituting $\epsilon_{\rm e}=0.1, \; p=2.17, \nu= 6 \;  {\rm GHz}, \; \theta_{\rm obs}=25^{\circ}, \theta_{\rm j}=5^{\circ}, D_{\rm L}=40$ Mpc, criterion 2 yields 
\begin{equation}
E_{\rm iso, 52}\gtrsim 0.4(n_{-4}\, \epsilon_{\rm B,-2})^{-0.79}.
\label{crit2}
\end{equation}

This condition is shown by the dot-dashed line in the bottom panel of Fig. \ref{paramspace}, where the different colors corresponds to different values of $\epsilon_{\rm B}=10^{-2},10^{-3},10^{-4}$ as indicated in the plot legend. The regions above the dot-dashed lines satisfy criterion 2 for the corresponding values of $\epsilon_{\rm B}$.

The final criterion for detectability requires that the flux at peak, $F_{\rm peak}$ is larger than the flux from the cocoon component. The latest observations show the afterglow of GW170817 has started to decline, and this decline follows a power law in time as roughly $\propto t^{-2}$ \citep{2018arXiv180502870A}. Attributing this emission to the cocoon, we estimate the late time flux from it by extrapolating this power to later times. For example, taking the 6~GHz measurements we can model the decline in flux as 
\begin{equation}
F_{\rm d}(t)\approx 0.018 \, t_{2.5}^{-2} \;{\rm mJy},
\end{equation}
with $t$  in days. This extrapolation is shown by the solid line in the top panel of Fig. \ref{paramspace}. Criterion 3 requires $F_{\rm peak}>F_{\rm d}(t_{\rm peak})$. Using the same parameter substitutions used to obtain equation \ref{crit2}, this condition yields

\begin{equation}
E_{\rm iso,52}\gtrsim\epsilon_{\rm B,-2}^{-0.476}\,n_{-4}^{-0.0755}.
\label{crit3}
\end{equation}
This equality is shown by the solid lines in the bottom panel of Fig. \ref{paramspace} where different colors correspond to different values of $\epsilon_{\rm B}$ color coded in the same way as for criterion 2 (the dot-dashed lines). Therefore the inequality \ref{crit3} is satisfied for regions above the solid lines.
So the regions in Fig. \ref{paramspace} which satisfy all three criteria must lie above the dashed black line (criterion 1), and above the dot-dashed and solid lines (criterion 2 and 3) corresponding to the same value of $\epsilon_{\rm B}$. These regions have been shaded for better visualization. In order for the second peak to be detectable, the parameters pertaining to the jet must lie in the shaded regions.

\subsection{Second peak  light curves using  the  ``Afterglow library''}
\label{library}

\begin{figure}
\includegraphics[width=\columnwidth]{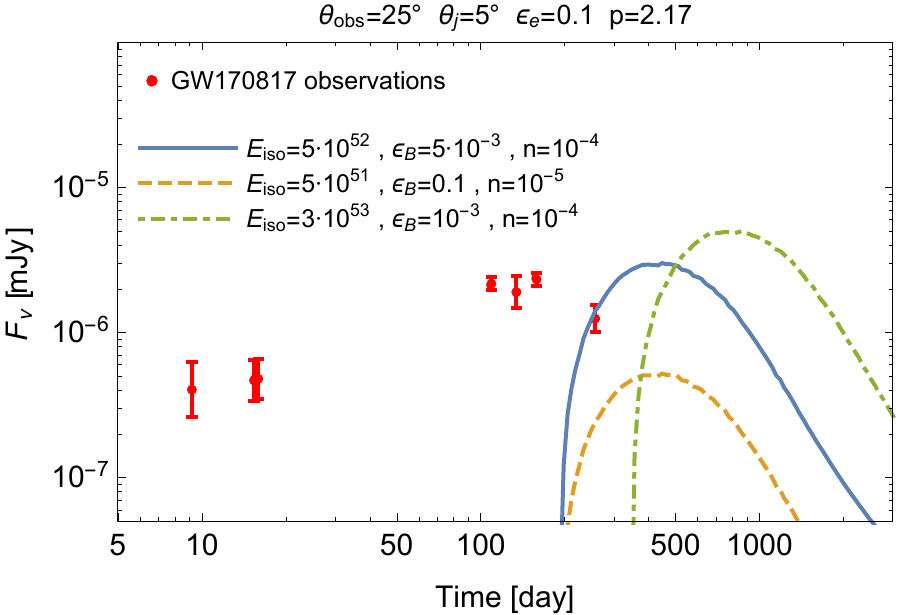} \\
\includegraphics[width=\columnwidth]{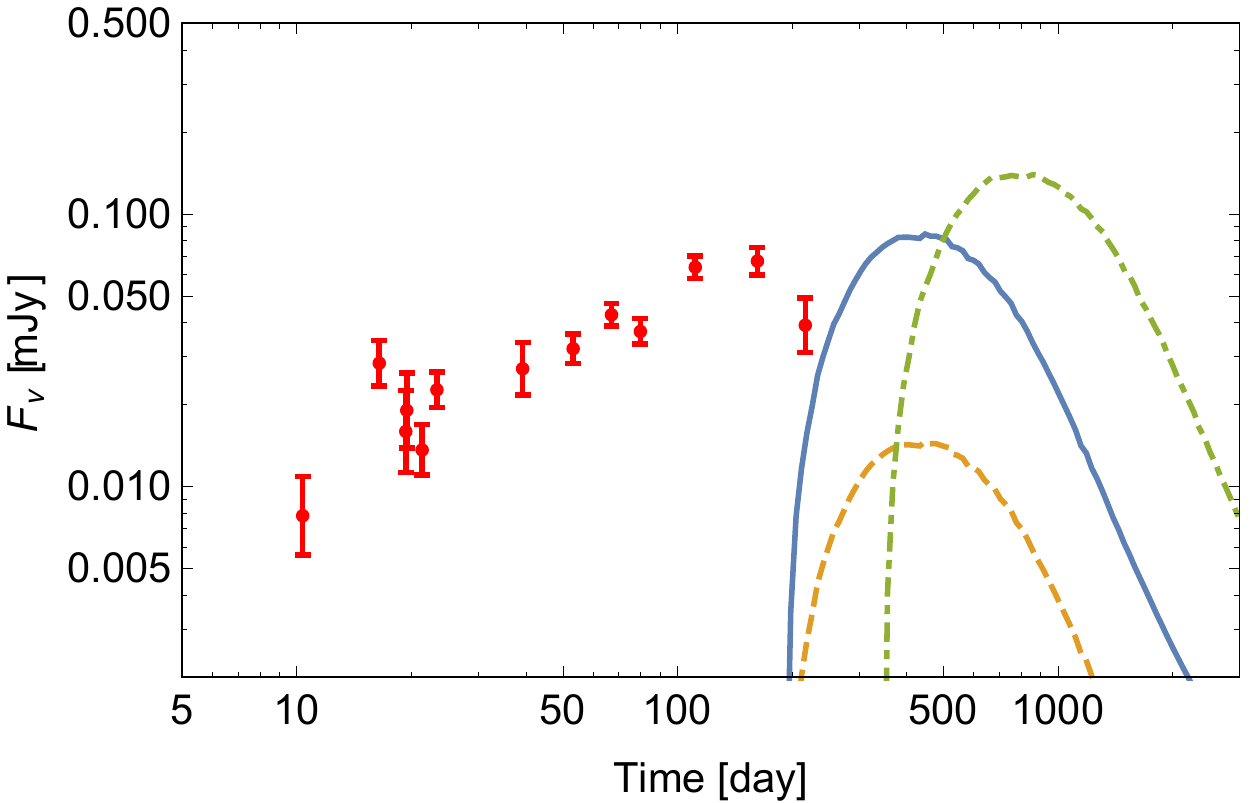}
\caption{Light curves in X-ray at 1 keV (Top) and Radio at 6 GHz (Bottom) using the  ``Afterglow library''. Red points show observations of GW170817 \citep{2018arXiv180103531M}, and lines show the afterglow from a jet calculated using the afterglow library \citep{vaneerten2012} for two sets of parameters (see sec.. for description of parameters). In a \fjc{} model, the earlier time observations (red points) can be attributed to emission from the cocoon. The afterglow from a jet will peak at later times for off-axis observers, in this scenario, this peak can cause a second bump in the overall X-ray light curve of GW170817. For parameters used here, the second peak occurs at $\sim$500 days. }
\label{fig:sb}
\end{figure}

In this Section we discuss the calculations of 
the afterglow light curves using the ``Afterglow library'' \citep{2012ApJ...747L..30V},  which uses linear radiative transfer to calculate synchrotron light curves and spectra. We use a few different sets of parameters when calculating the afterglow (maybe point to a table or plot listing the parameters). Guided by observations (references), we fix the spectral slope, $p=2.17$ and the observing angle w.r.t jet axis, $\theta_{\rm obs}=25^{\rm o}$. We use a typical value of $\epsilon_{\rm e} = 0.1$, which denotes the fraction of energy in the electrons of the shocked fluid, and a relatively narrow `top-hat' jet with opening angle $\theta_j$=5 deg. We vary the isotropic equivalent energy $E_{\rm iso}$, the fraction of energy in the magnetic field of the shocked fluid $\epsilon_{\rm B}$, and number density of external medium $n$ cm$^{-3}$. The resulting light curves for these different sets of parameters are shown in fig~\ref{fig:sb} the observations of GW170817 (taken from \cite{2018arXiv180103531M}) are also shown in the same plots for radio (6~GHz) and X-ray (1~keV). These plots demonstrate the possibility where the afterglow from the jet can cause a late time rise in the light curve of GW170817.

\subsection{Emission from off-axis forward shock: numerical calculation of  synchrotron emission}
\label{inhouse}
Next, we use the classic forward shock model of \cite{1996ApJ...473..204S,1998ApJ...497L..17S} to calculate the  emission seen by an off-axis observer. As a novel feature, we calculate numerically the radiative and cooling losses of the particles. To do so, we first find Greens function for particles injected at some moment in time into the forward shock, then we integrate over different injections times, and allowing for time-of-flight delays we calculated the  expected light curve.

We assume self-similar relativistic shock with $\Gamma \propto t^{-{m}/{2}} $ with  $m=3$  \citep{1976PhFl...19.1130B} 
\citep[thus, we neglect lateral evolution of the shock -  it is small, see][]{2010ApJ...722..235V,2012MNRAS.421..522L}. 
The  minimum Lorentz factor of accelerated electrons in the shock frame   is then \begin{eqnarray}
\gamma_{\rm min}^{\prime} \propto \epsilon_e (\Gamma-1) \frac{m_p}{m_e},
\end{eqnarray}
while the comoving magnetic field is 
\begin{eqnarray}
\frac{{B^{\prime}}^2}{8 \pi} = \epsilon_B n_p (\Gamma-1) \Gamma m_p c^2 \propto t^{-m} \propto {t^{\prime}}^{-\frac{2m}{2+m}}
\end{eqnarray}
(primed quantities are  in the  fluid frame and un-primed are in the coordinate frame.)
Thus, the magnetic field strength satisfies
\begin{eqnarray}
B^{\prime}=B_0^{\prime} \left(\frac{t^{\prime}}{t_0^{\prime}}\right)^{-\frac{m}{m+2}}
\label{newB}
\end{eqnarray}
where the magnetic field is $B_0$ and  Lorentz factor is $\Gamma=\Gamma_0$ at time $t^{\prime}=t_0^{\prime}$. 

Particles  are injection with distribution function $f_{inj}$ at time $t_i^{\prime}$ through an area $A$. 
The Lorentz factor of particles evolves according to
\ba&&
\frac{{d}\gamma^{\prime}}{{dt^{\prime}}}=-\frac{\tilde{C}_2 B_0^2 {\gamma^{\prime}}^2}{{t^{\prime}}^{\frac{2m}{2+m}}}-\frac{m \gamma^{\prime}}{2 t^{\prime} \left(2+m\right)}
\nn &&
\tilde{C}_2=\frac{\sigma _T {t_0^{\prime}}^{\frac{2m}{2+m}}}{6 \text{$\pi$$m_e$c}}
\label{newgammapde}
\ea
where the first term describes radiative losses and the second the adiabatic expansion.

The  evolution of the distribution function is  then described  by, first, solving  for the Greens function $G(\gamma^{\prime}, t^{\prime}, t_i^{\prime})$
\begin{gather}
\frac{\partial G(\gamma^{\prime},t^{\prime})}{\partial t^{\prime}} = \frac{\tilde{C}_2 B_0^2}{{t^{\prime}}^{\frac{2m}{2+m}}}\frac{\partial \left({\gamma^{\prime}}^2G(\gamma^{\prime},t^{\prime})\right)}{\partial \gamma^{\prime}} \nonumber +\frac{m}{2 t^{\prime} \left(2+m\right)}\frac{\partial (\gamma^{\prime} G(\gamma^{\prime},t^{\prime}))}{\partial \gamma^{\prime}}\\
+f_{inj}^{\prime}\left(\gamma^{\prime},t^{\prime}_i\right)  \delta (t^{\prime} - t_i^{\prime})
\label{newpde}
\end{gather}
with injection
\begin{eqnarray}
f_{inj}^{\prime}\left(\gamma^{\prime},t^{\prime}_i\right)=f_i^{\prime} {\gamma^{\prime}}^{-p} \Theta(\gamma^{\prime}- \gamma_{\rm min}^{\prime}(t_i^{\prime}))
\end{eqnarray}
where $f_i'$ satisfies $f_i' \int_{\gamma_{\rm min}^{\prime}}^{\infty} {\gamma^{\prime}}^{-p} d\gamma^{\prime}=nAc$.  And, second, integrating with the injection rate
\begin{eqnarray}
{f}(\gamma^{\prime},t^{\prime})=\int G(\gamma^{\prime},t^{\prime},t_i^{\prime}) f_i^{\prime} dt_i^{\prime},
\end{eqnarray}
where $f(\gamma^{\prime}, t^{\prime})$ is the total distribution.

Consider a jet (actually, a shock) with opening angle $\theta_j$  viewed at an angle $\theta_{\rm obs}$.  Emissivity at each moment is given by an integral over the shock surface
\begin{gather}
L^{\prime}(\omega^{\prime}, t^{\prime})=\int \int {\frac{f(\gamma^{\prime}, t^{\prime})}{A} P(\omega^{\prime})} \, d\gamma^{\prime} dA\nonumber \\
\approx \int_{\theta_{\rm obs}-\theta_j}^{\theta_{\rm obs}+\theta_j} \int_{\phi_{\rm min}\theta}^{\phi_{\rm max}\theta} \int _{\gamma_{\min }^{\prime}}^{\infty } \frac{r^{2}\sin\theta f(\gamma^{\prime} ,t^{\prime}) P(\omega^{\prime})}{2 \pi (c t^{\prime} \Gamma)^2 (1-\cos\theta_j)} d\gamma^{\prime} d\phi d\theta\nonumber \\
= \int_{\theta_{\rm obs}-\theta_j}^{\theta_{\rm obs}+\theta_j} \int _{\gamma_{\min }^{\prime}}^{\infty } \frac{(\phi_{\rm max}\theta-\phi_{\rm min}\theta) r^{2}\sin\theta f(\gamma^{\prime} ,t^{\prime}) P(\omega^{\prime})}{2 \pi (c t^{\prime} \Gamma)^2 (1-\cos\theta_j)} d\gamma^{\prime}  d\theta
\label{LL}
\end{gather}
where $f$ is the distribution function,  $P_\om$ is the synchrotron power per unit frequency emitted by each electron  and we assumed that  $\theta_{\rm obs}$ is larger than $\theta_j$

The photons emitted by different parts of the jet  at the same moment  will arrive at different time  due to time-of-flight effects.
The distance between the initial explosion point and an emission point $(r, \theta)$ is $r=\varv t(1-\beta\cos\theta)^{-1}$, so the surfaces that corresponds to the instantaneous emission have relation: 
\begin{eqnarray}
r=\varv t=\frac{\varv T_0}{1-\beta}=\frac{vT_{\theta_i}}{1-\beta\cos(\theta_i)}
\label{surface}
\end{eqnarray}
where $T_0$ represents the observe time at $\theta=0$, and $T_{\theta_i}$ represents the observe time at $\theta=\theta_i$.

The time $t$  is measured in lab frame; the  corresponding  observe time $T_{\rm obs}$ is a function of $\theta$: $t=\frac{T_{\rm obs}}{1-\beta\cos\theta}$. So 
\begin{eqnarray}
\sin\theta d\theta = - \frac{T_{\rm obs}}{{t}^2 \beta} dt \approx - \frac{T_{\rm obs}}{{t}^2} dt
\end{eqnarray}
The geometric relation between $\theta$ and $\phi$ is 
\begin{eqnarray}
\phi_{\rm max}\theta-\phi_{\rm min}\theta = 2 \arccos \left(\frac{\cos\theta_j-\cos(\theta_{\rm obs})\cos\theta}{\sin(\theta_{\rm obs})\sin\theta}\right)
\end{eqnarray} 
Using  $t=t^{\prime} \Gamma$, the equation (\ref{LL}) becomes
\begin{gather}
L^{\prime} \approx \int_{t_{\theta^{\prime}=\theta_{\rm obs}-\theta_j}^{\prime}}^{t_{\theta^{\prime}=\theta_{\rm obs}+\theta_j}^{\prime}} \int _{\gamma_{\min }^{\prime}}^{\infty } - 2 \arccos\left(\frac{\cos\theta_j-\cos(\theta_{\rm obs})\cos\theta}{\sin(\theta_{\rm obs})\sin\theta}\right)\times \nonumber\\ 
\frac{c^2 T_{ob} \Gamma f(\gamma^{\prime} ,t^{\prime}) P(\omega^{\prime})}{2 \pi (c t^{\prime} \Gamma)^2 (1-\cos\theta_j)} d\gamma^{\prime} dt^{\prime}
\label{drop}
\end{gather}

Taking into  account  Doppler boosting, we finally arrive at the equation for the observed spectral luminosity as function of the observer time
\begin{gather}
L \approx \int_{t_{\theta^{\prime}=\theta_{\rm obs}-\theta_j}^{\prime}}^{t_{\theta^{\prime}=\theta_{\rm obs}+\theta_j}^{\prime}} \int _{\gamma_{\min }^{\prime}}^{\infty } \arccos \left(\frac{\cos\theta_j-\cos(\theta_{\rm obs})\cos\theta}{\sin(\theta_{\rm obs})\sin\theta}\right) \times \nonumber \\ \frac{T_{\rm obs} \delta^3 P(\frac{\omega}{\delta}) \int G(\gamma^{\prime},t^{\prime},t_i^{\prime}) f_i^{\prime} dt_i^{\prime}}{\pi {t^{\prime}}^2 \Gamma (1-\cos\theta_j)} d\gamma^{\prime} dt^{\prime}
\label{final}
\end{gather}
where  
\be
T_{\rm obs} = \frac{ \left(2+m\right) }{ \left(1+m\right) }\frac{t^{\prime}}{\Gamma}
\ee

\begin{figure}
\includegraphics[width=\columnwidth]{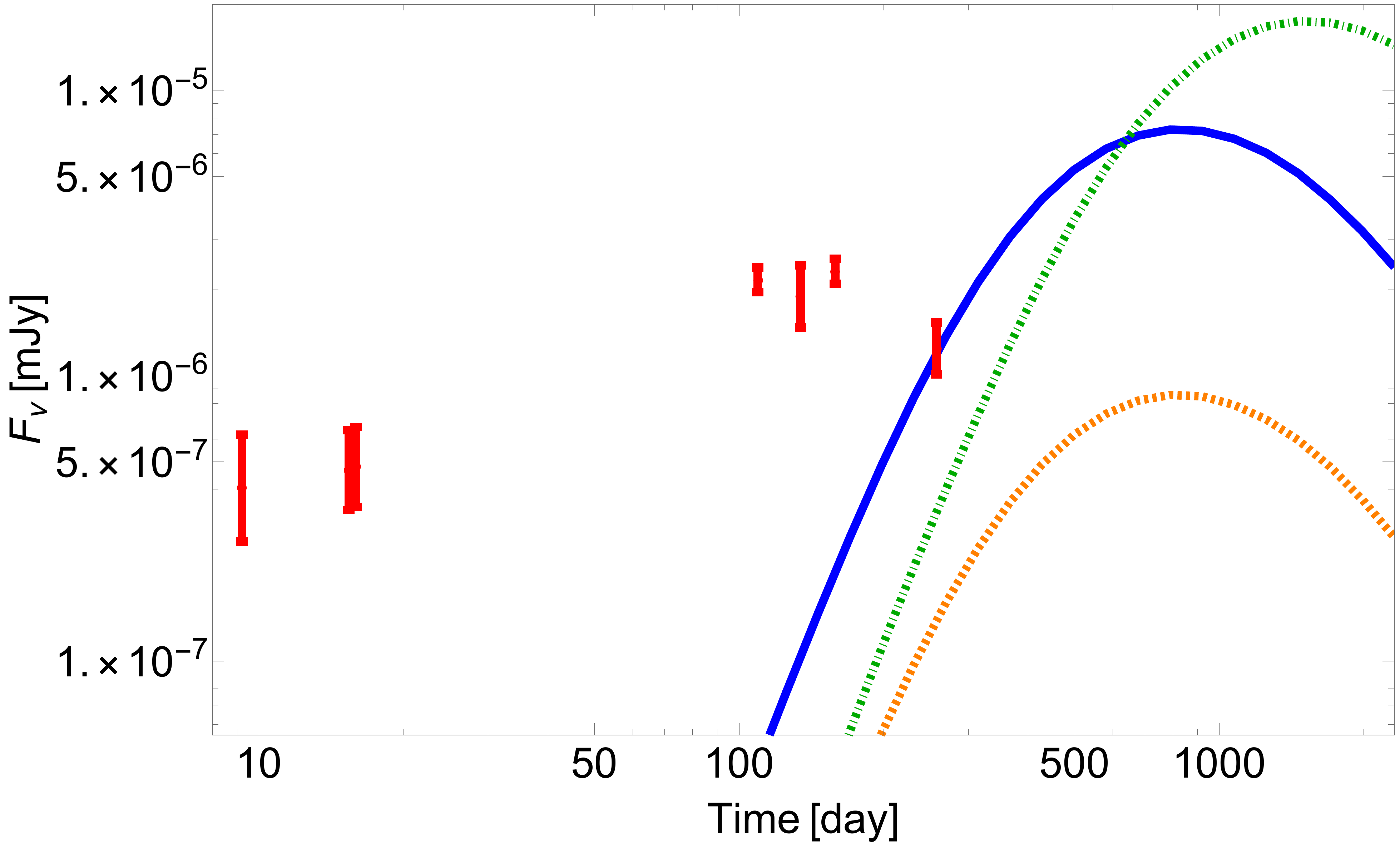}
\caption{Light curves in X-ray at 1 keV using the radiative calculations, see text for details. Red points show the observation of GW170817 at 1 KeV \citep{2018arXiv180502870A}. All light curves have the same parameter value as Figure 3.}
\label{second_peak_xray}
\end{figure}

Using the above procedure  we calculate the light curve behavior at 1 keV  for the same set of   parameters as in  Fig. \ref{fig:sb}, see Fig.\ref{second_peak_xray}. Within a factor of a few the two methods produce similar results.

\section{Conclusions}

In this paper we argue that  late afterglow of GRB/GW170817 may experience a second peak in brightness, as the fast primary jet, which avoided detection so far, becomes visible. Using three different approaches -- basic analytic estimates, ``Afterglow library'' and new radiative calculations --
we  put constrains on the jet energetics and microphysical parameters for the second peak to be observable.

Detectability of the second peak depends on macroscopic parameter (energy of the primary jet $E_{\rm iso}$ and external density $n$), as well as microscopic parameters ($\epsilon_e$ and $\epsilon_{\rm B} $), as well as the viewing angle. Observations of the prompt emission and the corresponding GRB constrain the viewing angle to be $\sim 20-30^\circ$ \citep[\eg][]{2017ApJ...848L..12A, 2018ApJ...852L..30P, 2018arXiv180502870A}.
Thus, we are left with $E_{\rm iso}$, $n$ and $\epsilon_e,\,\epsilon_B$. 

Our results indicate that even a mildly  energetic jet, with  $E_{\rm iso} \sim$ few $\times 10^{51}$ ergs may be detected even for low external density $n \sim 10^{-5}$ $cm^{-3}$ (if $\epsilon_e$ and $\epsilon_B$ are not too small). This compares favorably with the expected jet power (\ref{EISO}).

We end our conclusions with general remark: 
if the jet has more energy than the cocoon, it should show up as a second distinct peak.
This is because the other parameters ($n$, $\epsilon_{\rm e}$, and $\epsilon_{\rm B}$) should actually be the same for the 2 afterglow components. In the case of radiatively inefficient jet, the late time emission just tracks the total true energy of the jet.
We can even argue that since the jet drives the cocoon, it is typically true that $E_{\rm jet}>E_{\rm cocoon}$, i.e. we should see two bump structure on light curves from radio through  X-ray.

\section*{Acknowledgements}

This work had been supported by   NSF  grant AST-1306672, DoE grant DE-SC0016369 and NASA grant 80NSSC17K0757. DG and AK acknowledge support from NASA grants NNX16AB32G and NNX17AG21G.






\begin{thebibliography}{}
\makeatletter
\relax
\def\mn@urlcharsother{\let\do\@makeother \do\$\do\&\do\#\do\^\do\_\do\%\do\~}
\def\mn@doi{\begingroup\mn@urlcharsother \@ifnextchar [ {\mn@doi@}
  {\mn@doi@[]}}
\def\mn@doi@[#1]#2{\def\@tempa{#1}\ifx\@tempa\@empty \href
  {http://dx.doi.org/#2} {doi:#2}\else \href {http://dx.doi.org/#2} {#1}\fi
  \endgroup}
\def\mn@eprint#1#2{\mn@eprint@#1:#2::\@nil}
\def\mn@eprint@arXiv#1{\href {http://arxiv.org/abs/#1} {{\tt arXiv:#1}}}
\def\mn@eprint@dblp#1{\href {http://dblp.uni-trier.de/rec/bibtex/#1.xml}
  {dblp:#1}}
\def\mn@eprint@#1:#2:#3:#4\@nil{\def\@tempa {#1}\def\@tempb {#2}\def\@tempc
  {#3}\ifx \@tempc \@empty \let \@tempc \@tempb \let \@tempb \@tempa \fi \ifx
  \@tempb \@empty \def\@tempb {arXiv}\fi \@ifundefined
  {mn@eprint@\@tempb}{\@tempb:\@tempc}{\expandafter \expandafter \csname
  mn@eprint@\@tempb\endcsname \expandafter{\@tempc}}}

\bibitem[\protect\citeauthoryear{{Abbott} et~al.,}{{Abbott}
  et~al.}{2017a}]{2017ApJ...848L..12A}
{Abbott} B.~P.,  et~al., 2017a, \mn@doi [\apjl] {10.3847/2041-8213/aa91c9},
  \href {http://adsabs.harvard.edu/abs/2017ApJ...848L..12A} {848, L12}

\bibitem[\protect\citeauthoryear{{Abbott} et~al.,}{{Abbott}
  et~al.}{2017b}]{2017ApJ...848L..13A}
{Abbott} B.~P.,  et~al., 2017b, \mn@doi [\apjl] {10.3847/2041-8213/aa920c},
  \href {http://adsabs.harvard.edu/abs/2017ApJ...848L..13A} {848, L13}

\bibitem[\protect\citeauthoryear{{Alexander} et~al.,}{{Alexander}
  et~al.}{2018}]{2018arXiv180502870A}
{Alexander} K.~D.,  et~al., 2018, preprint, \href
  {http://adsabs.harvard.edu/abs/2018arXiv180502870A} {} (\mn@eprint {arXiv}
  {1805.02870})

\bibitem[\protect\citeauthoryear{{Barkov} \& {Komissarov}}{{Barkov} \&
  {Komissarov}}{2008}]{BK08b}
{Barkov} M.~V.,  {Komissarov} S.~S.,  2008, \mn@doi [\mnras]
  {10.1111/j.1745-3933.2008.00427.x}, \href
  {http://adsabs.harvard.edu/abs/2008MNRAS.385L..28B} {385, L28}

\bibitem[\protect\citeauthoryear{{Barkov} \& {Pozanenko}}{{Barkov} \&
  {Pozanenko}}{2011}]{bp11}
{Barkov} M.~V.,  {Pozanenko} A.~S.,  2011, \mn@doi [\mnras]
  {10.1111/j.1365-2966.2011.19398.x}, \href
  {http://adsabs.harvard.edu/abs/2011MNRAS.417.2161B} {417, 2161}

\bibitem[\protect\citeauthoryear{{Blandford} \& {Begelman}}{{Blandford} \&
  {Begelman}}{1999}]{bb99}
{Blandford} R.~D.,  {Begelman} M.~C.,  1999, \mn@doi [\mnras]
  {10.1046/j.1365-8711.1999.02358.x}, \href
  {http://adsabs.harvard.edu/abs/1999MNRAS.303L...1B} {303, L1}

\bibitem[\protect\citeauthoryear{{Blandford} \& {McKee}}{{Blandford} \&
  {McKee}}{1976}]{1976PhFl...19.1130B}
{Blandford} R.~D.,  {McKee} C.~F.,  1976, \mn@doi [Physics of Fluids]
  {10.1063/1.861619}, \href {http://adsabs.harvard.edu/abs/1976PhFl...19.1130B}
  {19, 1130}

\bibitem[\protect\citeauthoryear{{Blandford} \& {Znajek}}{{Blandford} \&
  {Znajek}}{1977}]{bz77}
{Blandford} R.~D.,  {Znajek} R.~L.,  1977, \mnras, \href
  {http://adsabs.harvard.edu/abs/1977MNRAS.179..433B} {179, 433}

\bibitem[\protect\citeauthoryear{{Blinnikov}, {Novikov}, {Perevodchikova}  \&
  {Polnarev}}{{Blinnikov} et~al.}{1984}]{bnpp84}
{Blinnikov} S.~I.,  {Novikov} I.~D.,  {Perevodchikova} T.~V.,   {Polnarev}
  A.~G.,  1984, Pis ma Astronomicheskii Zhurnal, \href
  {http://adsabs.harvard.edu/abs/1984PAZh...10..422B} {10, 422}

\bibitem[\protect\citeauthoryear{{Bromberg}, {Tchekhovskoy}, {Gottlieb},
  {Nakar}  \& {Piran}}{{Bromberg} et~al.}{2018}]{2018MNRAS.475.2971B}
{Bromberg} O.,  {Tchekhovskoy} A.,  {Gottlieb} O.,  {Nakar} E.,   {Piran} T.,
  2018, \mn@doi [\mnras] {10.1093/mnras/stx3316}, \href
  {http://adsabs.harvard.edu/abs/2018MNRAS.475.2971B} {475, 2971}

\bibitem[\protect\citeauthoryear{{D'Avanzo} et~al.,}{{D'Avanzo}
  et~al.}{2018}]{davanzo2018}
{D'Avanzo} P.,  et~al., 2018, preprint, \href
  {https://ui.adsabs.harvard.edu/#abs/2018arXiv180106164D} {p.
  arXiv:1801.06164} (\mn@eprint {arXiv} {1801.06164})

\bibitem[\protect\citeauthoryear{{Eichler}, {Livio}, {Piran}  \&
  {Schramm}}{{Eichler} et~al.}{1989}]{1989Natur.340..126E}
{Eichler} D.,  {Livio} M.,  {Piran} T.,   {Schramm} D.~N.,  1989, \mn@doi
  [\nat] {10.1038/340126a0}, \href
  {http://adsabs.harvard.edu/abs/1989Natur.340..126E} {340, 126}

\bibitem[\protect\citeauthoryear{{Finstad}, {De}, {Brown}, {Berger}  \&
  {Biwer}}{{Finstad} et~al.}{2018}]{2018arXiv180404179F}
{Finstad} D.,  {De} S.,  {Brown} D.~A.,  {Berger} E.,   {Biwer} C.~M.,  2018,
  preprint, \href {http://adsabs.harvard.edu/abs/2018arXiv180404179F} {}
  (\mn@eprint {arXiv} {1804.04179})

\bibitem[\protect\citeauthoryear{{Goldstein}, {Veres}, {Burns}  \& et
  al.}{{Goldstein} et~al.}{2017}]{gol17}
{Goldstein} A.,  {Veres} P.,  {Burns} E.,   et al. 2017, \mn@doi [\apjl]
  {10.3847/2041-8213/aa8f41}, \href
  {http://adsabs.harvard.edu/abs/2017ApJ...848L..14G} {848, L14}

\bibitem[\protect\citeauthoryear{{Gottlieb}, {Nakar}  \& {Piran}}{{Gottlieb}
  et~al.}{2018}]{gnp17}
{Gottlieb} O.,  {Nakar} E.,   {Piran} T.,  2018, \mn@doi [\mnras]
  {10.1093/mnras/stx2357}, \href
  {http://adsabs.harvard.edu/abs/2018MNRAS.473..576G} {473, 576}

\bibitem[\protect\citeauthoryear{{Granot}, {Gill}, {Guetta}  \& {De
  Colle}}{{Granot} et~al.}{2017}]{2017arXiv171006421G}
{Granot} J.,  {Gill} R.,  {Guetta} D.,   {De Colle} F.,  2017, preprint, \href
  {http://adsabs.harvard.edu/abs/2017arXiv171006421G} {} (\mn@eprint {arXiv}
  {1710.06421})

\bibitem[\protect\citeauthoryear{{Kasliwal} et~al.,}{{Kasliwal}
  et~al.}{2017}]{kasliwal2017}
{Kasliwal} M.~M.,  et~al., 2017, \mn@doi [Science] {10.1126/science.aap9455},
  \href {https://ui.adsabs.harvard.edu/#abs/2017Sci...358.1559K} {358, 1559}

\bibitem[\protect\citeauthoryear{{Kathirgamaraju}, {Barniol Duran}  \&
  {Giannios}}{{Kathirgamaraju} et~al.}{2017}]{kbg17}
{Kathirgamaraju} A.,  {Barniol Duran} R.,   {Giannios} D.,  2017,
  ArXiv:1708.07488, \href {http://adsabs.harvard.edu/abs/2017arXiv170807488K}
  {}

\bibitem[\protect\citeauthoryear{{LIGO Scientific Collaboration} \& {Virgo
  Collaboration}}{{LIGO Scientific Collaboration} \& {Virgo
  Collaboration}}{2017}]{GW-trigger}
{LIGO Scientific Collaboration} {Virgo Collaboration} 2017, LVC GRB Coordinates
  Network, 21505

\bibitem[\protect\citeauthoryear{{Lamb} \& {Kobayashi}}{{Lamb} \&
  {Kobayashi}}{2018}]{lamb2018}
{Lamb} G.~P.,  {Kobayashi} S.,  2018, \mn@doi [\mnras] {10.1093/mnras/sty1108},
  \href {https://ui.adsabs.harvard.edu/#abs/2018MNRAS.tmp.1056L} {p.~1056}

\bibitem[\protect\citeauthoryear{{Lazzati}, {Perna}, {Morsony},
  {L{\'o}pez-C{\'a}mara}, {Cantiello}, {Ciolfi}, {giacomazzo}  \&
  {Workman}}{{Lazzati} et~al.}{2017}]{lazzati2017}
{Lazzati} D.,  {Perna} R.,  {Morsony} B.~J.,  {L{\'o}pez-C{\'a}mara} D.,
  {Cantiello} M.,  {Ciolfi} R.,  {giacomazzo} B.,   {Workman} J.~C.,  2017,
  preprint, \href {https://ui.adsabs.harvard.edu/#abs/2017arXiv171203237L} {p.
  arXiv:1712.03237} (\mn@eprint {arXiv} {1712.03237})

\bibitem[\protect\citeauthoryear{{Lehner}, {Palenzuela}, {Liebling}, {Thompson}
   \& {Hanna}}{{Lehner} et~al.}{2012}]{2012PhRvD..86j4035L}
{Lehner} L.,  {Palenzuela} C.,  {Liebling} S.~L.,  {Thompson} C.,   {Hanna} C.,
   2012, \mn@doi [\prd] {10.1103/PhysRevD.86.104035}, \href
  {http://adsabs.harvard.edu/abs/2012PhRvD..86j4035L} {86, 104035}

\bibitem[\protect\citeauthoryear{{Lyutikov}}{{Lyutikov}}{2012}]{2012MNRAS.421..522L}
{Lyutikov} M.,  2012, \mn@doi [\mnras] {10.1111/j.1365-2966.2011.20331.x},
  \href {http://adsabs.harvard.edu/abs/2012MNRAS.421..522L} {421, 522}

\bibitem[\protect\citeauthoryear{{Margutti} et~al.,}{{Margutti}
  et~al.}{2018}]{2018arXiv180103531M}
{Margutti} R.,  et~al., 2018, ArXiv:1801.03531, \href
  {http://adsabs.harvard.edu/abs/2018arXiv180103531M} {}

\bibitem[\protect\citeauthoryear{{Metzger}, {Piro}  \& {Quataert}}{{Metzger}
  et~al.}{2008}]{mpq08}
{Metzger} B.~D.,  {Piro} A.~L.,   {Quataert} E.,  2008, \mn@doi [\mnras]
  {10.1111/j.1365-2966.2008.13789.x}, \href
  {http://adsabs.harvard.edu/abs/2008MNRAS.390..781M} {390, 781}

\bibitem[\protect\citeauthoryear{{Mooley} et~al.,}{{Mooley}
  et~al.}{2018}]{mooley2018}
{Mooley} K.~P.,  et~al., 2018, \mn@doi [\nat] {10.1038/nature25452}, \href
  {https://ui.adsabs.harvard.edu/#abs/2018Natur.554..207M} {554, 207}

\bibitem[\protect\citeauthoryear{{Nakar}, {Piran}  \& {Granot}}{{Nakar}
  et~al.}{2002}]{2002ApJ...579..699N}
{Nakar} E.,  {Piran} T.,   {Granot} J.,  2002, \mn@doi [\apj] {10.1086/342791},
  \href {http://adsabs.harvard.edu/abs/2002ApJ...579..699N} {579, 699}

\bibitem[\protect\citeauthoryear{{Paczynski}}{{Paczynski}}{1986}]{p86}
{Paczynski} B.,  1986, \mn@doi [\apjl] {10.1086/184740}, \href
  {http://adsabs.harvard.edu/abs/1986ApJ...308L..43P} {308, L43}

\bibitem[\protect\citeauthoryear{{Perego}, {Radice}  \& {Bernuzzi}}{{Perego}
  et~al.}{2017}]{2017ApJ...850L..37P}
{Perego} A.,  {Radice} D.,   {Bernuzzi} S.,  2017, \mn@doi [\apjl]
  {10.3847/2041-8213/aa9ab9}, \href
  {http://adsabs.harvard.edu/abs/2017ApJ...850L..37P} {850, L37}

\bibitem[\protect\citeauthoryear{{Pozanenko} et~al.,}{{Pozanenko}
  et~al.}{2018}]{2018ApJ...852L..30P}
{Pozanenko} A.~S.,  et~al., 2018, \mn@doi [\apjl] {10.3847/2041-8213/aaa2f6},
  \href {http://adsabs.harvard.edu/abs/2018ApJ...852L..30P} {852, L30}

\bibitem[\protect\citeauthoryear{{Radice}, {Galeazzi}, {Lippuner}, {Roberts},
  {Ott}  \& {Rezzolla}}{{Radice} et~al.}{2016}]{rgl16}
{Radice} D.,  {Galeazzi} F.,  {Lippuner} J.,  {Roberts} L.~F.,  {Ott} C.~D.,
  {Rezzolla} L.,  2016, \mn@doi [\mnras] {10.1093/mnras/stw1227}, \href
  {http://adsabs.harvard.edu/abs/2016MNRAS.460.3255R} {460, 3255}

\bibitem[\protect\citeauthoryear{{Rezzolla}, {Giacomazzo}, {Baiotti}, {Granot},
  {Kouveliotou}  \& {Aloy}}{{Rezzolla} et~al.}{2011}]{2011ApJ...732L...6R}
{Rezzolla} L.,  {Giacomazzo} B.,  {Baiotti} L.,  {Granot} J.,  {Kouveliotou}
  C.,   {Aloy} M.~A.,  2011, \mn@doi [\apjl] {10.1088/2041-8205/732/1/L6},
  \href {http://adsabs.harvard.edu/abs/2011ApJ...732L...6R} {732, L6}

\bibitem[\protect\citeauthoryear{{Ruiz}, {Lang}, {Paschalidis}  \&
  {Shapiro}}{{Ruiz} et~al.}{2016}]{rlps16}
{Ruiz} M.,  {Lang} R.~N.,  {Paschalidis} V.,   {Shapiro} S.~L.,  2016, \mn@doi
  [\apjl] {10.3847/2041-8205/824/1/L6}, \href
  {http://adsabs.harvard.edu/abs/2016ApJ...824L...6R} {824, L6}

\bibitem[\protect\citeauthoryear{{Sari}, {Narayan}  \& {Piran}}{{Sari}
  et~al.}{1996}]{1996ApJ...473..204S}
{Sari} R.,  {Narayan} R.,   {Piran} T.,  1996, \mn@doi [\apj] {10.1086/178136},
  \href {http://adsabs.harvard.edu/abs/1996ApJ...473..204S} {473, 204}

\bibitem[\protect\citeauthoryear{{Sari}, {Piran}  \& {Narayan}}{{Sari}
  et~al.}{1998}]{1998ApJ...497L..17S}
{Sari} R.,  {Piran} T.,   {Narayan} R.,  1998, \mn@doi [\apjl]
  {10.1086/311269}, \href {http://adsabs.harvard.edu/abs/1998ApJ...497L..17S}
  {497, L17}

\bibitem[\protect\citeauthoryear{{Savchenko}, {Ferrigno}, {Kuulkers}  \& et
  al.}{{Savchenko} et~al.}{2017}]{sav17}
{Savchenko} V.,  {Ferrigno} C.,  {Kuulkers} E.,   et al. 2017, \mn@doi [\apjl]
  {10.3847/2041-8213/aa8f94}, \href
  {http://adsabs.harvard.edu/abs/2017ApJ...848L..15S} {848, L15}

\bibitem[\protect\citeauthoryear{{Troja} et~al.,}{{Troja}
  et~al.}{2017}]{2017Natur.551...71T}
{Troja} E.,  et~al., 2017, \mn@doi [\nat] {10.1038/nature24290}, \href
  {http://adsabs.harvard.edu/abs/2017Natur.551...71T} {551, 71}

\bibitem[\protect\citeauthoryear{{Troja} et~al.,}{{Troja}
  et~al.}{2018a}]{2018arXiv180106516T}
{Troja} E.,  et~al., 2018a, ArXiv:1801.06516, \href
  {http://adsabs.harvard.edu/abs/2018arXiv180106516T} {}

\bibitem[\protect\citeauthoryear{{Troja} et~al.,}{{Troja}
  et~al.}{2018b}]{troja2018}
{Troja} E.,  et~al., 2018b, \mn@doi [\mnras] {10.1093/mnrasl/sly061}, \href
  {https://ui.adsabs.harvard.edu/#abs/2018MNRAS.tmpL..60T} {p.~L60}

\bibitem[\protect\citeauthoryear{{Xie}, {Zrake}  \& {MacFadyen}}{{Xie}
  et~al.}{2018}]{xie2018}
{Xie} X.,  {Zrake} J.,   {MacFadyen} A.,  2018, preprint, \href
  {https://ui.adsabs.harvard.edu/#abs/2018arXiv180409345X} {p.
  arXiv:1804.09345} (\mn@eprint {arXiv} {1804.09345})

\bibitem[\protect\citeauthoryear{{van Eerten} \& {MacFadyen}}{{van Eerten} \&
  {MacFadyen}}{2012a}]{2012ApJ...747L..30V}
{van Eerten} H.~J.,  {MacFadyen} A.~I.,  2012a, \mn@doi [\apjl]
  {10.1088/2041-8205/747/2/L30}, \href
  {http://adsabs.harvard.edu/abs/2012ApJ...747L..30V} {747, L30}

\bibitem[\protect\citeauthoryear{{van Eerten} \& {MacFadyen}}{{van Eerten} \&
  {MacFadyen}}{2012b}]{vaneerten2012}
{van Eerten} H.~J.,  {MacFadyen} A.~I.,  2012b, \mn@doi [\apj]
  {10.1088/0004-637X/751/2/155}, \href
  {http://adsabs.harvard.edu/abs/2012ApJ...751..155V} {751, 155}

\bibitem[\protect\citeauthoryear{{van Eerten}, {Zhang}  \& {MacFadyen}}{{van
  Eerten} et~al.}{2010}]{2010ApJ...722..235V}
{van Eerten} H.,  {Zhang} W.,   {MacFadyen} A.,  2010, \mn@doi [\apj]
  {10.1088/0004-637X/722/1/235}, \href
  {http://adsabs.harvard.edu/abs/2010ApJ...722..235V} {722, 235}

\makeatother
\end{thebibliography}






\bsp	
\label{lastpage}
\end{document}